\documentstyle[prd,aps]{revtex}

\begin{document}  
\draft
\twocolumn[\hsize\textwidth\columnwidth\hsize\csname
@twocolumnfalse\endcsname

\title{Reconciling Supersymmetry and Left-Right Symmetry}

\author{Charanjit S. Aulakh$^{(1)}$, Karim Benakli$^{(2)}$ and 
Goran Senjanovi\'c$^{(3)}$}

\address{$^{(1)}$ {\it Dept. of Physics, Panjab University, 
Chandigarh, India   }}
\address{$^{(2)}$ {\it Phys. Dept.,  Texas A \& M, College Station,
 USA. }}
\address{$^{(3)}${\it International Center for Theoretical Physics,
 Trieste, Italy }}

\maketitle

\begin{abstract}
We construct the minimal supersymmetric left-right theory
and show that at the renormalizable level it
requires the existence of an intermediate $B-L$ breaking scale. The 
subsequent symmetry breaking down to MSSM automatically preserves
 R-symmetry.
Furthermore, unlike in the nonsupersymmetric version of the theory, 
the see-saw mechanism takes its canonical form. The theory predicts 
the existence of a triplet of Higgs scalars much lighter
 than the $B-L$ breaking scale.

\end{abstract}\vskip1pc]

{\bf A. Introduction.}\hspace{0.5cm} There is no doubt that the 
Minimal Supersymmetric Standard Model (MSSM)
 has become 
the most popular extension of the Standard Model (SM).  However, 
one of the most  appealing 
features of the Standard Model is lost in its supersymmetric
 counterpart: automatic conservation of baryon and lepton numbers. 
In SUSY, unless some mechanism of suppression is 
found, baryon number violation, as is well known, is
 catastrophically fast.

It turns out that another popular extension of the Standard Model,
 the Left-Right (L-R) symmetric theory \cite{ps74} offers a 
natural solution to this MSSM problem. The B-L  symmetry, which is
 a part of L-R models, automatically forbids all the baryon and 
lepton number violating operators\cite{moh}.  L-R theories are 
interesting in their own right, for among other appealing features, 
they offer a simple and natural
 explanation of the smallness of neutrino mass 
through the so-called see-saw mechanism \cite{seesaw,ms80}.

In view of this, it becomes important to systematically study 
 L-R supersymmetric theories, in order to arrive at  a
realistic  minimal supersymmetric left-right model (MSLRM). Up
 to now, the only serious attempt in this direction is the work 
of Kuchimanchi and Mohapatra \cite{km94} which showed that in 
the minimal version of the theory no spontaneous symmetry 
breaking takes place \cite{helsinki}. Furthermore, when this 
is cured through the
 introduction of a parity-odd singlet, the soft SUSY breaking
 terms inevitably lead to the breaking of electromagnetic charge
 invariance.
 This is true at least for a scale of L-R symmetry breaking $M_R $
 above $10 TeV$. In this letter we stick to the physically
 motivated assumption of $M_R$ being much larger than the scale 
of supersymmetry breaking $M_S$ taken to be not far from  the
 electroweak scale: $M_S \simeq M_W$. We show that this problem
 disappears if one allows for an intermediate $B-L$ 
 breaking scale.  Furthermore, the
 physically unappealing singlet becomes redundant.

The most important result of our study is that {\em at  
low energies the model reduces to the MSSM with an exact 
R-parity}: its breaking  is simply incompatible with phenomenology. 
 A phenomenologically interesting 
feature of the theory is the possibility of a low-lying $B-L$ 
scale, $M_{BL}\stackrel{>}{\sim} 1 TeV$ .

Furthermore, the see-saw mechanism in this theory takes 
its canonical form $ m_\nu \simeq m_D^2/M_{BL} $ (where $m_D$ 
is the neutrino Dirac mass term), as opposed to the 
nonsupersymmetric version of L-R models or $SO(10)$ GUTs.
 Namely, despite its generic see-saw form, the  neutrino mass 
in ordinary L-R theories depends unfortunately 
on the unknown parameters of the Higgs potential.  

Another important prediction of the theory regards the Higgs masses: 
one finds an $SU(2)_L$ triplet with a mass of the order of 
$M_{BL}^2/M_R$ (or $M_S$, depending which scale is bigger). This 
could provide a crucial test for the theory with low $M_{BL}$.

{\bf B. The minimal model: a brief review.} \hspace{0.5cm}
For the sake of self consistency, and in order to pave the way
 for the realistic model, we first review briefly the minimal
 model.

The so-called minimal supersymmetric left-right model is based
 on the gauge group $SU(3)_c \times
SU(2)_L \times SU(2)_R \times U(1)_{B-L}$. It contains three
 generations of quark and leptonic chiral superfields with the 
following transformation properties:
\begin{eqnarray}
Q=(3,2,1,1/3)&\;\;\;\;\;  & Q_c=(3^*,1,2,-1/3) \nonumber \\ 
L=(1,2,1,-1)& \;\;\;\;\; & L_c=(1,1,2,1) 
\end{eqnarray}
where the numbers in the brackets denote the quantum numbers under
$SU(3)_c$, $SU(2)_L$, $ SU(2)_R$ and $ U(1)_{B-L}$ respectively.
The Higgs sector consists of 
\begin{eqnarray}
 \Phi_i=(1,2,2,0) \quad (i = 1, 2)
 \nonumber \\
  \Delta=(1,3,1,2)  , \quad 
  \bar{\Delta} =(1,3,1,-2) \nonumber \\
\Delta_c=(1,1,3,-2), \quad \bar{\Delta_c} =(1,1,3,2)
\end{eqnarray}
The number of bidoublets is doubled in order to achieve a
 nonvanishing CKM quark mixing matrix, and the number of triplets
 is doubled for the sake of anomaly cancellations. 

The gauge symmetry is augmented by a discrete parity or left-right 
symmetry under which the fields transform as
\begin{eqnarray}
Q              \leftrightarrow     {Q_c}^* ,\quad 
L              \leftrightarrow     {L_c}^* ,\quad
\Phi_i         \leftrightarrow     {\Phi_i}^\dagger,  \nonumber \\
\Delta         \leftrightarrow  {\Delta_c}^* ,\quad
\bar{\Delta}   \leftrightarrow   \bar{\Delta_c}^*.
\end{eqnarray}

The minimal model suffers from an incurable disease: {\it it 
cannot break parity spontaneously} \cite{km94}.
 One possible way out is to
add a parity-odd
 singlet \cite{c85} which in our opinion is not so appealing.
 Moreover, although now parity could be spontaneously broken, it
 turns out that the same happens to the electromagnetic charge. 

In this theory, as   
 Kuchimanchi and Mohapatra show \cite{km94}, the vacuum manifold
contains a circle   parametrized by an angle $\theta$
\begin{equation}
 \left < \Delta^{c} \right > 
= d \left ( \begin{array}{cc} 0 & sin \theta \\
                     cos \theta & 0  \end{array} \right ), \quad
\left < \bar\Delta_c \right > = \bar d 
  \left ( \begin{array}{cc} 0 & cos \theta \\
                     sin \theta & 0 \end{array}\right )
\label{deltavev2}
\end{equation}
where $d=\bar d$ in the absence of soft SUSY breaking terms.
The problem appears when these terms are switched on, since in 
general the soft mass terms for $\Delta_c$ and $\bar\Delta_c$ 
will be different, whereas left-right symmetry was forcing them 
to be equal in the original superpotential valid at the scale of 
parity breaking $M_R$. 
In other words, at the scale of SUSY breaking $M_S$ the world is not 
left-right symmetric anymore. 
Thus $d=\bar d$ no longer holds, and
it can be shown  that the minimum corresponds to $\theta = \pi/4$,
which  breaks electromagnetic charge invariance.
Notice that there is no hope that we live in the false 
charge-preserving vacuum, due to the original continuous 
degeneracy. Our vacuum falls classically
(without need for quantum tunneling) into the true charge-breaking
 one. 

To avoid this, one could resort to the  use nonrenormalizable 
higher-dimensional terms as suggested in \cite{mr96}. We
 prefer in what follows to focus on the phenomenologically 
attractive possibility of an intermediate $B-L$ 
breaking scale.

{\bf D. The B-L route. } \hspace{0.5cm}
The idea here, often discussed in the context of ordinary 
L-R models,   is to break $SU(2)_R$ down to its subgroup 
$U(1)_R$ while preserving $B-L$. This is achieved by
including two new Higgs superfields $\Omega$  and $\Omega_c$ with 
the following quantum numbers \cite{km94}

\begin{equation}
\Omega = (1,3,1,0) , \quad \Omega_c = (1,1,3,0)
\end{equation}
where under parity $\Omega \to \Omega_c^*$.

{\em What is new however is the fact that there is no need for the
 parity-odd singlet $\Sigma$.} This in our opinion is an 
important result and it tells us that in a sense this model 
is a realistic MSLRM at least at the renormalizable level.
Furthermore, the vev of the triplet $\Omega_c$ splits the masses of the
$SU(2)_L\times U(1)$ Higgs doublets in the bidoublets $\Phi$, allowing
thus for  the MSSM at low energies. 

We now show that parity can be broken spontaneously and at the same 
time electromagnetic charge is automatically preserved. The 
effect of introducing the B-L neutral triplets $\Omega, \Omega_c$
is best appreciated by first considering the extremization of the 
potential at  high scales $M_R \gg M_S,M_W$, where the effect of 
the soft breaking terms  is negligible so that the potential has
 the form it takes for a supersymmetric gauge theory with
 superpotential  
\begin{eqnarray}
 W_{LR}&=& {\bf h}_l^{(i)} L^T \tau_2 \Phi_i \tau_2 L_c 
+ {\bf h}_q^{(i)} Q^T \tau_2 \Phi_i \tau_2 Q_c 
       +i {\bf f} L^T \tau_2 \Delta L\nonumber \\    
     &  &+i {\bf f}^* L^{cT}\tau_2 \Delta_c L_c + 
m_\Delta  {\rm  Tr}\, \Delta \bar{\Delta} 
       + m_\Delta^*  {\rm Tr}\,\Delta_c \bar{\Delta_c} \nonumber \\
   & & +{m_{\Omega} \over 2}  {\rm Tr}\,\Omega^2
          +{m_{\Omega}^* \over 2}  {\rm Tr}\,\Omega_c^2
 + \mu_{ij} {\rm Tr}\,  \tau_2 \Phi^T_i \tau_2 \Phi_j 
        \nonumber \\
  & &     +a {\rm Tr}\,\Delta \Omega \bar{\Delta}
       +a^* {\rm Tr}\,\Delta_c \Omega_c \bar{\Delta_c} \nonumber \\
& &  + \alpha_{ij} {\rm Tr}\, \Omega  \Phi_i \tau_2 \Phi_j^T \tau_2 
       +\alpha_{ij}^* {\rm Tr}\, \Omega_c  \Phi^T_i \tau_2 \Phi_j 
\tau_2 
\label{superpot}
\end{eqnarray}
with ${\bf h}^{(i)}_{q,l}  =  {{\bf h}^{(i)}_{q,l}}^\dagger $, 
$\mu_{ij}  =  \mu_{ji} = \mu_{ij}^*$,
$\alpha_{ij} = -\alpha_{ji}$, ${\bf f}$ and ${\bf h}$ are symmetric
 matrices,  and generation and color indices are understood.

Supersymmetry implies  F-flatness conditions given by the following
equations for the scalar fields.
\begin{eqnarray}
F_{\bar\Delta} &=& m_\Delta \Delta +
 a (\Delta \Omega -{1\over 2} {\rm Tr}\,\Delta\Omega) =0 \nonumber 
 \\
F_{\bar{\Delta_c}} &=& m_\Delta^*  \Delta_c  
+  a^* (\Delta_c \Omega_c -{1\over 2} {\rm Tr}\,\Delta_c\Omega_c)=0
\nonumber  \\ 
F_\Delta &=& m_\Delta  \bar\Delta +i {\bf f} L L^T\tau_2 
+ a (\Omega\bar\Delta-{1\over 2}{\rm Tr}\,\Omega\bar\Delta)=0
\nonumber  \\
F_{\Delta_c} &=& m_\Delta^* \bar\Delta_c +i {\bf f^*}
  L_c L_c^T \tau_2
+ a^* (\Omega_c{\bar\Delta_c}-{1\over 2}{\rm Tr}\,
\Omega_c\bar\Delta_c)
=0\nonumber  \\
F_\Omega &=& m_\Omega \Omega
       +a (\bar\Delta\Delta  -{1\over 2} {\rm Tr}\,\bar\Delta\Delta)
= 0 \nonumber \\    
F_{\Omega_c} &=& m_{\Omega}^* \Omega_c
       +a^* (\bar\Delta_c\Delta_c  -{1\over 2} {\rm Tr}\,\bar\Delta_c
 \Delta_c)
=0\nonumber \\
F_{L} &=& 2 i {\bf f} \tau_2 \Delta L = 0\nonumber \\
F_{L_c} &=& 2 i {\bf f^*} \tau_2 \Delta_c L_c = 0
\label{fflat}
\end{eqnarray}
In the above drop the $\Phi$ fields, which must have zero vevs at $M_R$.
 It is easy to show that $\langle \Phi \rangle =0$ is consistent
 with (\ref{fflat})

We also have to satisfy the  D-flat conditions, namely

\begin{eqnarray}
D_{R i} &= & 2 {\rm Tr}\,\Delta_c^\dagger\tau_i\Delta_c + 2 
{\rm Tr}\,\bar\Delta_c^\dagger\tau_i\bar \Delta_c\nonumber \\  
& & + 2 {\rm Tr}\,\Omega_c^\dagger\tau_i\Omega_c
+ L_c^\dagger \tau_i L_c = 0 \nonumber \\
D_{L i} &=&  2 {\rm Tr}\,\Delta^{ \dagger}\tau_i\Delta 
+ 2 {\rm Tr}\,\bar\Delta^{ \dagger}\tau_i\bar \Delta \nonumber \\
& & +  2 {\rm Tr}\,\Omega^{ \dagger}\tau_i\Omega + L^{\dagger} \tau_i
 L 
 = 0 \nonumber  \\
D_{B-L} &= & -L^\dagger L  + 2 {\rm Tr}\,(\Delta^{ \dagger}\Delta 
- \bar\Delta^{ \dagger}\bar \Delta)\nonumber \\
& & +L_c^\dagger L_c   - 2 {\rm Tr}\,(\Delta_c^\dagger\Delta_c 
- \bar\Delta_c^\dagger\bar \Delta_c )=0
\label{dflat}
\end{eqnarray}

Here we keep the left-handed fields 
since we have to show that parity can be broken spontaneously 
and at the same time we wish to know whether R-parity is broken or
 not.

Typically in SUSY theories minimization of the D-term potential 
(in our case equation (\ref{dflat})) leads 
to a number of flat directions which may be characterized 
 by the set of holomorphic gauge invariants formed 
from the chiral multiplets \cite{lt96}. Then, one uses the 
vanishing  of the F-potential
(\ref{fflat}) in an attempt to determine as much as possible of these
 holomorphic functions. One can use this elegant method to prove 
that in this theory a parity-broken minimum leads to a  
determination of these gauge invariants, therefore lifting  the 
flat directions (again, neglecting the squarks fields as in the
 MSSM). Due to the lack of space, we leave this analysis for a 
separate publication, and instead present here a straightforward
 analysis that leads to the determination of the vacuum manifold.

It is obvious from (\ref{fflat}) and (\ref{dflat}) that  the
left-handed vevs can be taken to vanish
\begin{equation}
\langle \Delta \rangle = \langle \bar\Delta \rangle =
\langle \Omega \rangle =
\langle L \rangle = 0
\end{equation}

We should mention that in this case $\langle \Phi \rangle$ {\em must} 
vanish, as can be easily seen by minimizing $V_F$ and $V_D$.
Although clearly there is a solution in which
 the right-handed counterpart fields also have vanishing vevs,
 and no symmetry is broken, we now focus on the realistic 
parity-breaking case.

First notice that  (\ref{fflat}) gives 
\begin{equation}
{\rm Tr}\,\Delta^2_c = {\rm Tr}\,\Delta_c \Omega_c=0.
\label{ceroinv}
\end{equation}
By an appropriate $SU(2)_R$ rotation one may put $\Delta_c$ in the
 form 

\begin{equation}
\left < \Delta_c \right > =\left ( \matrix{
                      0 &\langle \delta_c^{--}\rangle \cr
                     \langle\delta_c^0\rangle &0 \cr } \right )
\label{rotation}
\end{equation}

where superscripts denote electromagnetic charges 
\begin{equation}
Q_{em}= T_{3L} + T_{3R}+
{{B-L}\over 2} .
\end{equation}

Now (\ref{ceroinv}) gives $\langle\delta_c^{--}\rangle\langle 
\delta_c^0\rangle = 0$, which implies the electromagnetic
 charge-preserving form for  $\langle \Delta_c \rangle$. Next, 
from $F_{L_c}=0$, baring accidental
 cancellations among different families; and using again
 (\ref{ceroinv}) and  $ D_{B-L} - D_{3 R} =0$, it is an easy 
exercise to show that $\langle L_c\rangle $ vanishes, and that
$\langle \Omega_c\rangle $ and $\langle \bar\Delta_c \rangle$
preserve $Q_{em}$. In short

\begin{eqnarray}
 \langle \Omega_c \rangle= \left ( \begin{array}{cc} w & 0 \\
                     0 & -w  \end{array} \right ), \quad
 \langle L_c  \rangle = 0 \nonumber \\
\langle \Delta_c \rangle = 
  \left ( \begin{array}{cc} 0 & 0 \\
                     d & 0 \end{array}\right ),\quad
\langle \bar\Delta_c \rangle  =  
  \left ( \begin{array}{cc} 0 & \bar d \\
                     0 & 0 \end{array}\right )
\label{vev1}
\end{eqnarray} 

This proves the two important claims we made earlier. First, 
that the electromagnetic charge invariance of this vacuum is 
automatic for any value of the parameters of the theory
 (of course, neglecting as we did the squarks fields). Second, 
that the symmetry breaking 
down to the MSSM preserves R-parity since 
$\left< L \right>=\left< L_c \right>=0$ generation by generation. 
Of course, as often happens in supersymmetry, this vacuum is 
degenerate with the unbroken one. The important point is that 
now they are not connected continuously.

With the remaining D- and F-equations it is straightforward to
 find the absolute values of the nonvanishing vevs

\begin{equation}
| w | = \left | {m_\Delta \over a}\right |
\equiv M_R ,\quad
|d| = |\bar d|= \left |{m_\Delta m_\Omega \over
a^2}\right |^{1/2} \equiv M_{BL}
\label{vev2}
\end{equation}

Notice an interesting property of (\ref{vev2}). If we wish to have 
$M_R \gg M_{BL}$, 
 we need $m_\Delta\gg m_\Omega$, i.e. a sort of inverse hierarchy 
of the mass scales. The same situation is encountered in the case 
of the P-odd singlet.

The analysis of the Higgs mass spectrum proceeds as usual, with expected
results, except for the mass  of the $\Omega$ triplet. Instead of being
$M_{BL}$ as one would imagine naively, it turns out to be of order
$M_{BL}^2/M_R$.

{\bf E. Low energy effective theory and R-parity.} \hspace{0.5cm}
An important question that must be faced is what happens when 
the soft supersymmetry breaking terms are turned on. Specifically,
 one would like to know the fate of R-parity. In order to answer
 this question we need to have an effective low-energy theory after
 the heavy fields are integrated out. 

We are not interested here in the small corrections $M_S$ to the
 large vevs of order $M_R$, but it is crucial to know whether 
the vevs we have found to vanish can be turned on after SUSY
 breaking.  In the case of heavy fields, heavy meaning having a
 mass of order $M_R$, this can only happen if they have linear
 couplings to fields that either already have a vev or are 
light. By light fields we mean the ones that acquire masses only
after   SUSY and $SU(2)\times U(1)$ breaking, and which thus are
 allowed to get vevs at that stage. The relevant fields are
 obviously $L, L_c$ and the bidoublets $\Phi_i$;  $L_c$ being
 heavy and the others light.

 It is easy to check that due to the trilinear terms in the
 supersymmetric potential $ L_c$  can get a vev only if $L$ 
acquires one, and we have $\langle L_c\rangle \simeq \langle
 L\rangle  M_S/M_R$. Thus there is no R-parity violation in the 
right-handed sector until after it is broken (if at all) by the
 vev of the left-handed sneutrino.

We show now that phenomenological considerations prevent  this from 
happening. Notice first that in the limit of infinite $M_R$, the 
MSLRM reduces to the MSSM with an exact R-parity.
Namely, when $\Omega_c$ gets a vev, the couplings $\alpha $ in Eq.
(\ref{superpot}) lead to the splitting of the bidoublets into two light
$SU(2)\times U(1)$ doublets and two heavy ones (with masses proportional
to $\langle \omega_c\rangle$). Of course, the light doublets are light
only with  the usual fine-tuning 
between the $\mu$ and the $\alpha(\Omega_c)$ terms in the effective
potential.
 
 In this
case $\langle L
 \rangle \neq 0$ \cite{am82} would imply the existence of the Majoron
 \cite{cmp81}, which corresponds to the spontaneous breaking of the 
global B-L symmetry. Such a Majoron can be ruled out due to its 
couplings to the Z-boson.

Next, for finite $M_R$, it is a simple excercise to show that the 
Majoron becomes massive and, as expected on general grounds, one finds

\begin{equation}
m_J^2 \simeq {M_S^3 \over M_R}
\end{equation}
where $M_J$ is the Majoron's mass.  This follows from soft terms in 
the potential of the type

\begin{equation}
\Delta V_{\rm soft} = ... + M_S L^T \tau_2 \Phi_i \tau_2 L_c + ...
\end{equation}

Clearly for $M_R\gg M_S$ there is no possibility  that $M_J >
 M_{Z}$, and the bounds on the doublet Majoron from the $Z$
 width in fact rule out the possibility that $\langle \tilde\nu
 \rangle \neq 0$.

Thus we have a remarkable prediction: {\em the low-energy
 effective theory of the MSLRM is the MSSM with unbroken R-parity,
 and the lightest supersymmetric partner (LSP) is stable}. This
 has profound phenomenological and specially cosmological 
consequences. In particular it allows the LSP to be a dark
 matter candidate.

{\bf F. See-saw mechanism} \hspace{0.5cm} Maybe the nicest 
feature of the theory is the implementation of the see-saw 
mechanism. As is well known, in the ordinary L-R symmetric
 theories, the left-handed triplet $\Delta$ necessarily gets a
 nonvanishing vacuum-expectation value \cite{ms80} 
\begin{equation}
\langle \Delta\rangle = \alpha {M_W^2 \over M_{BL}}
\end{equation}
where $\alpha$ is the ratio of some unknown couplings in the
 Higgs potential. This, while preserving the see-saw effect,
 unfortunately introduces additional
 unknown parameters and spoils the  canonical form we cited in 
the introduction. However, in the supersymmetric version, as we
 have seen, $\Delta $ has no  vev due to the absence of tadpole
 terms  in the effective Higgs  potential. Thus the see-saw 
mechanism is ``clean'', since it only depends on the neutrino
 Dirac mass terms, i.e.
\begin{equation}
m_\nu \simeq {m_D^2\over M_{BL}}
\end{equation}

 This is especially important when
 one studies the $SO(10)$ extensions of these theories, where
 the Dirac neutrino masses became related to the up quark masses,
 and the see-saw mechanism becomes potentially predictive once
 the intermediate mass scale $M_{BL}$ is determined.

{\bf G. Summary and Outlook.}\hspace{0.5cm} 

Supersymmetry and left-right symmetry have grown with time  
into the central extension of the standard model, and L-R
 symmetry seems to play an important role in providing a gauge 
rationale for R-parity. However, a construction of the SUSY L-R
 theory is by no means trivial. As we know from 
the work of Kuchimanchi and Mohapatra \cite{km94}, and as we have
 reviewed here, in the minimal version of the theory the 
symmetry cannot be spontaneously  broken, whereas  when this 
is cured by the addition of a parity-odd singlet one ends up 
breaking also electromagnetic charge invariance.

The minimal price to be paid at the renormalizable level is 
then to accept an intermediate $B-L$ scale. Phenomenologically 
this of course is a blessing, for it leads to a whole plethora 
of new Higgs particles, potentially accessible to experiment. 
Of particular interest is the
triplet $\Omega$, whose mass is of order $max[M_{BL}^2/M_R ; M_S]$.

To summarize, we have found that the  intermediate $B-L$ scale 
not only solves the problem of charge-breaking minima, but also 
leads to the important phenomenological and cosmological
 prediction of a conserved R-parity. The gauge constraints of
 the theory simply do not 
allow for spontaneous breaking of R-parity, even after the soft SUSY
 breaking terms are turned on. Equally important, the see-saw
 mechanism takes its canonical form which in the context of GUTs
 such as, say, $SO(10)$,would lead to predictions for neutrino 
masses.

\vspace{0.5cm}
 We are deeply grateful to Alejandra Melfo and Andrija Ra\v{s}in
for many useful 
discussions, for their invaluable help in checking most of our 
computations and for the careful reading of our manuscript. 
We also acknowledge valuable discussions with Kaladi Babu 
and Anjan Joshipura. C.S.A. wishes to thank 
ICTP High Energy Group for the hospitality during the 
initial stages of this work. The work of K.B. is supported by a 
DOE grant DE-FGO3-95ER40917, and that of G.S. by EEC under the TMR
contract ERBFMRX-CT960090.


\begin{references}


\bibitem{ps74}J.C. Pati and A. Salam, Phys. Rev. {\bf D10}, 275 
             (1974); 
              R.N. Mohapatra and J.C. Pati, {\it ibid} {\bf D11},
               566; 2558 (1975); G. Senjanovi\'{c} and R.N.
              Mohapatra, {\it ibid} {\bf D12}, 1502 (1975).

\bibitem{moh} R.N. Mohapatra, Phys. Rev. {\bf D 34}, 3457 (1986).
\bibitem{seesaw}M. Gell-Mann, P. Ramond and R. Slansky, in {\it
Supergravity}, eds. P. van Niewenhuizen and D.Z. Freedman (North
Holland 1979); T. Yanagida, in Proceedings of {\it Workshop on
 Unified Theory and Baryon number in the Universe}, eds. 
O. Sawada and A. Sugamoto (KEK 1979);  R. N. Mohapatra and 
G. Senjanovi{\'c}, Phys. Rev. Lett. {\bf 44} (1980) 912.

\bibitem{ms80}  R. N. Mohapatra and G. Senjanovi{\'c}, Phys. Rev.
              {\bf D23}, 165 (1981).
\bibitem{km94}R. Kuchimanchi and R.N. Mohapatra, Phys. Rev. 
             {\bf D48},
              4352 (1993); Phys. Rev. Lett. {\bf 75}, 3989 (1995).
\bibitem{helsinki} We should mention here the work of the
Helsinki group,
see for example K. Huitu, M. Raidal and  J. Maalampi, in 
{\em Warsaw 1994, Proceedings, Physics from Planck scale to 
electroweak scale} 259-262, hep-ph/9412218; however their approach
 is manifestly L-R asymmetric, for they simply do not take into
 account the left-handed scalar fields. Thus they avoid the 
central issue of L-R symmetric theories, i.e. that of breaking
 parity spontaneously in a realistic theory.
\bibitem{c85} M. Cveti\v{c}, Phys. Lett. {\bf 164B}  55 (1985).

\bibitem{mr96}R.N. Mohapatra and A. Ra\v{s}in, 
            Phys. Rev.  {\bf D54 }, 5835 (1996);

            
\bibitem{lt96}For a rigorous treatment of holomorphic invariants
 corresponding to D-flat directions and for original references,
 see M.A. Luty and W. Taylor, Phys.Rev. {\bf D53}, 3399 (1996).

\bibitem{am82}C. Aulakh and R.N. Mohapatra, Phys. Lett. {\bf B119},
            136 (1982).
\bibitem{cmp81}Y. Chikashige, R.N. Mohapatra and R. Peccei, 
            Phys. Lett. {\bf B98}, 265 (1981). 


\end{references}
\end{document}